\documentclass[twocolumn,aps,showpacs]{revtex4}
\usepackage{graphicx}
\begin{document}
\title{Subsonic and supersonic nucleus-acoustic  solitary waves in thermally degenerate plasmas with heavy nucleus species}
 \author{A. A.  Mamun\footnote{Also at Wazed Miah Science Research Centre,
Jahangirnagar University, Savar, Dhaka-1342, Bangladesh.}} 
 \affiliation{Department of Physics,
Jahangirnagar University, Savar, Dhaka-1342, Bangladesh}

\begin{abstract}
A fully ionized multi-nucleus plasma system (containing thermally degenerate electron species, non-degenerate warm light nucleus species, and low dense stationary heavy nucleus species) is considered.  The basic features of thermal and degenerate pressure driven small and arbitrary amplitude subsonic and supersonic nucleus-acoustic  solitary waves in such a plasma system are studied by the pseudo-potential approach. The effects of stationary heavy nucleus,  non-relativistically and ultra-relativistically  electron degeneracies,  and  light nucleus temperature on small and arbitrary amplitude subsonic and supersonic  nucleus-acoustic solitary waves  are also examined.  It is found that (i) the presence of  stationary heavy nucleus species with Bolttzmann distributed electron species supports the existence of subsonic nucleus-acoustic solitary waves, and that the effects of electron degeneracies and  light nucleus temperature reduce the possibility for the formation of these subsonic nucleus-acoustic solitary waves; (ii) the amplitude (width) of the subsonic nucleus-acoustic solitary waves  increases (decreases) with the rise of the number density of heavy nucleus species;  (iii) the amplitude of the supersonic nucleus-acoustic solitary waves   in  the situation of  no-relativistically degenerate electron species  is much  smaller than that of ultra-relativistically degenerate electron species, but is much larger than that of isothermal electron species; (iv) their width in the situation of non-relativistically degenerate electron species is much wider than that of ultra-relativistically degenerate electron species; (v) their amplitude (width)  decreases (increases) with the rise of the light nucleus temperature.  The applications of the results in astrophysical, space, and laboratory plasma situations are briefly discussed. 
\end{abstract}
\pacs{52.35.Sb; 52.35.Mw; 52.35.Dm}
\date{\today}
\maketitle
\section{Introduction}
The ion-acoustic (IA)  waves \cite{Tonks29ia-theor,Revans33ia-expt,Buti68ia-lin} are thermal pressure driven longitudinal electro-acoustic waves in which the ion mass density (electron thermal pressure) provides the inertia (restoring force). They propagate as compression  and rarefaction (and vice-versa) of the inertial ion fluid  in  a plasma (a macroscopically neutral substance containing many interacting charged and neutral particles, which exhibit collective behavior due to the long-range Coulomb forces). The linear dispersion relation for the  IA waves propagating in a pure electron-ion plasma [containing Boltzmann distributed electron species (BDES) and cold inertial ion fluid]  is given by 
\begin{eqnarray}
\omega=\frac{kC_i}{\sqrt{1+k^2\lambda_{De}^2}},
\label{IA-dispersion}
\end{eqnarray}
where $\omega=2\pi f$ and $k=2\pi/\lambda$ in which $f$ ($\lambda$) is the IA wave frequency (wavelength);  $C_i=(k_BT_e/m_i)^{1/2}$ is  the IA speed in which $k_B$ is the Boltzmann constant, $T_e$ is the electron temperature, and $m_i$ is the ion mass; $\lambda_{De}=C_i/\omega_{pi}$ is the IA wave length scale in which  $\omega_{pi}=(4\pi N_{i0}e^2/m_i)^{1/2}$ is the ion plasma frequency, $N_{i0}$ is the ion number density at equilibrium, and $e$ is the magnitude of an electron charge. We note that for a pure electron-ion plasma $N_{e0}=N_{i0}$, where $N_{e0}$ is the electron number density at equilibrium. The dispersion relation (\ref{IA-dispersion}) indicates that for a long-wavelength limit (viz. $k\lambda_{De}\ll 1$), we have 
$\omega\simeq kC_i$, and for a  short wavelength limit  (viz. $k\lambda_{De}\gg 1$),  we have $\omega\simeq\omega_{pi}$.  So,  the upper limit of 
$\omega$ for the IA waves is  $\omega_{pi}$. 

The IA waves do not exit for cold electron limit ($T_e=0$).  However, in degenerate plasma (which is significantly  different from usual electron-ion plasma because of its extra-ordinarily high density \cite{Fowler1926,Chandrasekhar1931,Horn1991,Fowler1994,Koester2002,Shukla2011a,Shukla2011b,Brodin2016}, and low temperature \cite{Killian2006,Fletcher2006,Glenzer2009,Drake2009,Drake2010}), Mamun \cite{Mamun2018} predicted  the existence of degenerated pressure (generated due to Heisenberg's uncertainty principle with infinitesimally small uncertainty in position and infinitely large uncertainty in  momenta of degenerate electron species) driven nucleus-acoustic (NA) waves in absolutely cold degenerate plasmas.  The linear dispersion relation for such degenerate  pressure driven (DPD) NA waves in such a cold degenerate electron-nucleus plasma is given by  
\begin{eqnarray}
\omega=\frac{\sqrt{\gamma_e}kC_l}{\sqrt{1+ \gamma_ek^2\lambda_q^2}},
\label{NA-dispersion}
\end{eqnarray}
where $C_l= (Z_l{\cal E}_{e0}/m_l)^{1/2}$ is the DPD NA speed in which  ${\cal E}_{e0}=KN_{e0}^{\gamma_e-1}$ is the cold electron degenerate energy associated with the degenerate electron pressure \cite{Mamun2018}, $Z_l$ is  the charge state of the nucleus species, $m_l$ is the mass of the nucleus species, $K=3\pi\hbar^2/5m_e$ (where $m_e$ is the mass of an electron and  $\hbar$ is the reduced Plank constant) for $\gamma_e=5/3$ (non-relativistically degenerate electron species \cite{Shukla2011a,Shukla2011b}),  $K=3\hbar c/4$ (where $c$ is speed of light in vacuum) for $\gamma_e=4/3$ (ultra-relativistically degenerate electron species \cite{Shukla2011a,Shukla2011b}), and $K$ is unknown for $\gamma_e=1$ (which, thus, cannot be considered in  (\ref{NA-dispersion}), $N_{e0}$ is the degenerate electron number density at equilibrium; 
$\lambda_q=C_l/\omega_{pl}$ is the DPD NA wave length scale;  $\omega_{pl}= (4\pi N_{l0}Z_l^2e^2/m_l)^{1/2}$ is the nucleus plasma frequency. We note that for a degenerate electron-nucleus plasma $N_{e0}=Z_lN_{l0}$, where $N_{l0}$ is the non-degenerate nucleus number density at equilibrium.  The dispersion relation (\ref{NA-dispersion}) indicates that for the appropriate (long wavelength) limit, viz. $k\lambda_{D}\ll 1$, we have $\omega\simeq kC_l$, which indicates that in DPD NA waves, the nucleus mass density (electron degenerate pressure) provides the inertia (restoring force). We note that the DPD NA waves  disappear in absence of the electron degenerate pressure, which is independent of temperature of any plasma species. 

The dispersion relations (\ref{IA-dispersion}) and (\ref{NA-dispersion}) along with their interpretations indicate that the DPD NA waves  \cite{Mamun2018} are different from the IA waves \cite{Tonks29ia-theor,Revans33ia-expt,Buti68ia-lin}, and are  completely new  in view of restoring force, and are found to exist only in degenerate plasmas.  There are large  number of investigations  on the properties of linear and nonlinear  ion-acoustic, nucleus-acoustic, electron-acoustic, and positron-acoustic waves in degenerate plasma systems under different situations carried out during last one decade or more \cite{Mamun10a,Mamun10b,El-Labany10,Hossain11,Misra11,Akhtar11,Roy12,El-Taibany12a,El-Taibany12b,Haider12,Nahar13,Zobaer13a,Zobaer13b,El-Labany14,Rahman15,Hossen15,Hossen16,El-Labany16,MAS2016,MAS2017,Hasan17,Islam17a,Islam17b,Hosen17,Hosen18,Chowdhury18,Karmakar18,Das19,Fahad19,Patidar20}.
The investigations (mentioned above and made so far) on  IA or NA solitary waves (SWs) in degenerate plasmas (DPs), which are of our present interest, are based  absolutely cold degenerate plasma approximation and reductive perturbation method \cite{Washimi66}, and thus, are not valid for $T_e\ne0$ and $T_l\ne 0$ (where $T_l$ is the light nucleus temperature), and are also not valid for arbitrary amplitude SWs. 
So, (\ref{IA-dispersion}) is not valid when  the degenerate pressure  of the electron species is comparable to or greater than its thermal pressure, and that  (\ref{NA-dispersion}) is not valid when  the electron thermal pressure  is comparable to or greater than its degenerate pressure. We, therefore, propose a more general  degenerate plasma model by considering a thermally degenerate plasma (TDP) system containing thermally degenerate electron species (TDES), non-degenerate cold/warm light nucleus specie (since the degeneracy in nucleus species is at least 
$ (m_e/m_l)^2$ times lower than that in electron species \cite{Shukla2011b,Mamun2018,MAS2016,MAS2017}), and stationary heavy nucleus species (SHNS).  The advantage of this TDP model is that the TDES is valid for the arbitrary value of $\gamma_e$, i.e. valid for $\gamma_e=1$  (BDES),  $\gamma_e=5/3$ (non-relativistically TDES), $\gamma_e=4/3$ (ultra-relativistically TDES), etc.  The linear dispersion relation for the NA waves in such a TDP system with the cold NDLNS ($T_l=0$)  is given by
\begin{eqnarray}
\omega=\sqrt{\frac{\gamma_e}{1+\mu}}\frac{k C_q}{\sqrt{1+\frac{\gamma_e}{1+\mu} k^2\Lambda_q^2}},
\label{dis}
\end{eqnarray}
where $\mu=Z_hN_{h0}/Z_lN_{l0}$ with $Z_h$ ($N_{h0}$) is the charge state (number density) of the SHNS;  $C_q= (Z_l{\cal E}_{eT}/m_l)^{1/2}$, ${\cal E}_{eT}={\cal E}_{ed}+{\cal E}_{et}$, ${\cal E}_{ed}=KN_{e0}^{\gamma_e-1}$ is the energy associated with the electron degenerate pressure,   
${\cal E}_{et}=k_BT_eN_{e0}^{\gamma_e-1}$ is that associated with the electron thermal pressure, and  
$\Lambda_q=C_q/\omega_{pl}$. We note that for  BDES ($\gamma_e=1$), ${\cal E}_{eT}=k_BT_e$, $C_q=C_i$, 
and $\Lambda_q=\lambda_{De}$. So, in absence of SHNS ($\mu=0$),  (\ref{dis}) reduces to (\ref{IA-dispersion}). 

On the other hand, for the cold degenerate electron species ($T_e=0$),  ${\cal E}_{eT}={\cal E}_{e0}$, $C_q=C_l$, and $\Lambda_q=\lambda_q$.  So, in absence of the SHNS ($\mu=0$),  (\ref{dis}) reduces to (\ref{NA-dispersion}).  The dispersion relation (\ref{dis}) indicates that for the appropriate (long wavelength) limit,  viz. $k\Lambda_q\ll 1$, we have $\omega\simeq \sqrt{\gamma_e}kC_q/\sqrt{1+\mu}$, which indicates that in the NA waves, the light nucleus mass density (sum of electron degenerate and thermal  pressures) provides the inertia (restoring force), and  that the phase speed of the NA waves decreases (increases) with the rise of the value of $\mu$ ($\gamma_e$).  We note that the upper limit of  $\omega$ for the NA waves defined by (\ref{NA-dispersion}) and (\ref{dis}) is  $\omega_{pl}$. 

To the best knowledge of the author, no investigation has been made  on NA SWs  in any TDP system. Therefore, in the  present work, this new TDP model  is considered to investigate the arbitrary amplitude subsonic and supersonic SWs associated  the linear waves defined by (\ref{dis}).  The pseudo-potential approach \cite{Bernstein57,Cairns95}, which is valid for the arbitrary amplitude SWs, is used. 
The new TDP model proposed in the present work  is so general that it is  valid  not only for hot white dwarfs  \cite{Dufour08,Dufour11,Werner15,Werner19,Koester20},  but also for many space \cite{Rosenberg95,Havnes96,Tsintikidis96,Gelinas98} and laboratory  \cite{Fortov96,Fortov98,Mohideen98} plasma situations, where non-degenerate electron-ion plasma with heavy positively charged particles (as impurity or dust) occur.  

The manuscript is organized as follows. The governing equations in dimensional and normalized forms are given in Sec. II.  The conditions for the formation of subsonic and supersonic NA SWs associated with non-degenerate light nucleus species (NDLNS) in both cold  and warm adiabatic situations are described in Sec. III.  Their basic features are also illustrated in  the same section (Sec. III).  The discussion in short is provided in Sec. IV. 
\section{Model Equations}
We consider a TDP system containing TDES, NDLNS, and SHNS.  Thus, at equilibrium we have  $N_{e0}=Z_lN_{l0}+Z_hN_{h0}$. We also consider the propagation of the NA waves in  such a TDP system.  The dynamics of nonlinear NA waves in such a TDP system is described by 
\begin{eqnarray}
&&\frac{\partial N_j}{\partial T} +\frac{\partial}{\partial X}(N_jU_j) = 0,
\label{be1}\\
&&\frac{\partial {\cal P}_jq}{\partial T} +U_j\frac{\partial {\cal P}_jq}{\partial X} +\gamma_j{\cal P}_{jq}\frac{\partial U_j}{\partial X} = 0,
\label{be2}\\
&&\frac{\partial}{\partial X}({\cal P}_{ed}+{\cal P}_{et})-N_e e\frac{\partial\Phi}{\partial X}=0.
\label{be3}\\
&&\hat{D}_T U_l=-\frac{Z_le}{m_l}\frac{\partial\Phi}{\partial X}-\frac{1}{N_lm_l}\frac{\partial}{\partial X}({\cal P}_{ld}+{\cal P}_{lt}),
\label{be4}\\
&&\frac{\partial^2\Phi}{\partial X^2}=4\pi e(N_e-Z_lN_l- Z_lN_{h0}),
\label{be5}
\end{eqnarray}
where $\hat{D}_T={\partial}/{\partial T} +U_l{\partial}/{\partial X}$;  $\Phi$ is the electrostatic NA wave potential;  $N_j$ ($U_j$) is  number density (fluid speed) of the plasma species $j$ (with $j=e$ for the TDES, $j=l$ for the NDLNS);  ${\cal P}_{jq}$  in (\ref{be2})  and  (\ref{be4}) is the outward pressure for the species $j$ of the type $q$ (with $q=d$ for the degenerate pressure, and $q=t$ for the thermal pressure); $\gamma_j$ is the adiabatic index for the plasma species $j$; $X$ ($T$) is the space (time) variable.  The nonlinear equations (\ref{be1})$-$(\ref{be5}) describing  the nonlinear propagation of the NA waves in  the TDP system under consideration can be interpreted as follows:
\begin{itemize}
\item{Equation (\ref{be1}) is the continuity equation for the plasma species $j$, where the effects of the source and sink terms have been neglected.}

\item{Equation (\ref{be2}) is the energy equation for the plasma species $j$ for arbitrary $\gamma_j$. The use of this equation is meaningful if and only if the temperature $T_{j}$ is not constant since for constant 
$T_{j}$ , i.e. for $T_j=T_{j0}$ (where $T_{j0}$ is the temperature of the plasma species $j$ at equilibrium)  and 
$\gamma_j=1$ (BDES),  (\ref{be1}) and (\ref{be2}) are identical.}

\item{Equation (\ref{be3}) is the momentum balance equation for TDES, and is due to the fact that the sum of degenerate and thermal pressures (${\cal P}_{ed}+{\cal P}_{et}$) of the electron species counterbalances the electrostatic pressure ($N_ee\Phi$) associated with the NA waves. It is valid for $\omega/k\ll ({\cal E}_{eq}/m_e)^{1/2}$, where  $E_{eq}$ is the enegy associated with the 
electron thermal ($q=t$) or degenerate ($q=d$) pressure 
at equilibrium.}

\item{Equation (\ref{be4}) is the momentum balance equation for the NDLNS.  The last  term of (\ref{be4}) 
is due to the effect of the  sum of degenerate and thermal pressures (${\cal P}_{ld}+{\cal P}_{lt}$) of the NDLNS.}

\item{Equation (\ref{be5}) is Poisson's equation for the NA wave potential, which has closed the set of our basic equations  (\ref{be1})$-$ (\ref{be4}). The consideration of heavy nucleus species being stationary is valid since heavy nucleus plasma frequency is much less then the NA wave frequency because of the heavy mass and low number density of the heavy nucleus species.}
\end{itemize}
It is important to mention that the gravitational force acting in TDES and NDLNS, which is inherently very small compared to the other forces under consideration, is neglected for the study of the NA waves.

To find the expressions for ${\cal P}_{jq}$,  we start with  (\ref{be1}) and (\ref{be2}) and assume that all dependent variables in them depend on a single variable $\zeta= X-MT$ (with $M$ being the phase speed of the DA waves), and that $\partial /\partial T \rightarrow 0$. These assumptions allow us to express(\ref{be1}) and (\ref{be2})  as
\begin{eqnarray}
&&M\frac{dN_j}{d\zeta}-\frac{d}{d\zeta}(N_jU_j) = 0,
\label{P1}\\
&&M\frac{d{\cal  P}_{jq}}{d\zeta} -U_j\frac{d{\cal P}_{jq}}{d\zeta}-\gamma_j {\cal P}_j \frac{dU_j}{d\zeta}= 0.
\label{P2}
\end{eqnarray}
At equilibrium $N_j=N_{j0}$ and $U_j=0$.  Thus,  (\ref{P1}) can be expressed as
\begin{eqnarray}
&&U_j=M\left(1-\frac{N_{j0}}{N_j}\right)
\label{P3}
\end{eqnarray}
Now, substituting (\ref{P3}) into (\ref{P2}), and dividing the resulting equation by $N_j^{\gamma_j-1}$,  we obtain
\begin{eqnarray}
&&\frac{1}{N_j^{\gamma_j}}\frac{d{\cal P}_{jq}}{d\zeta} - \frac{\gamma_j{\cal P}_{jq}}{N_j^{(\gamma_j+1)}}\frac{dN_j}{d\zeta}= 0,
\label{P4}
\end{eqnarray}
which, after rearrangement, can be expressed as 
\begin{eqnarray}
&&\frac{d}{d\zeta}\left(\frac{{\cal P}_{jq}}{N_j^{\gamma_j}}\right)= 0.
\label{P5}
\end{eqnarray}
The integration of (\ref{P5}) yields
\begin{eqnarray}
{\cal P}_{jq}=K_{jq}N_j^{\gamma_j},
\label{P6}
\end{eqnarray}
where $K_{jq}={\cal E}_{jq}N_{j0}^{1-\gamma_j}$ is the integration constant [in which ${\cal E}_{jq}$ is equilibrium energy associated with the outward pressure for the species $j$ of type $q$.  

We now substitute $P_{ed}$ and $P_{et}$ [which can be obtained from (\ref{P6})] into (\ref{be3}),  and express  $n_e$ ($=N_e/N_{e0}$)  in terms of 
$\phi$ ($=e\Phi/{\cal E}_{eT}$, where  ${\cal E}_{eT}={\cal E}_{ed}+{\cal E}_{et}$) as
\begin{eqnarray}
n_e&=&\left(1+\frac{\gamma_e-1}{\gamma_e}\phi\right)^{\frac{1}{\gamma_e-1}}.
\label{ne}
\end{eqnarray}
It is important to note that (\ref{ne}) is valid for arbitrary value of $\gamma_e$, and is, thus, valid for non-relativistically ($\gamma_e=5/3$)  as well as ultra-relativistically ($\gamma_e=4/3$) TDES.   We note that for cold DES,  ${\cal E}_{et}=0$ and ${\cal E}_{eT}={\cal E}_{ed}=K_{ed}N_{e0}^{\gamma_e-1}$, which mean that $\phi=e\Phi/{\cal E}_{ed}$.  On the other hand,  for BDES,  
${\cal E}_{ed}=0$ and ${\cal E}_{eT}={\cal E}_{et}=k_BT_e$, which indicate that $\phi=e\Phi/k_BT_e$ and $\sigma_l=T_i/Z_lT_e$. 
 It is also important to mention that we cannot put $\gamma_e=1$ [corresponding to BDES, $n_e=\exp(\phi)$ with $\phi=e\Phi/k_BT_e$] directly into (\ref{ne}). To consider this limit,  we  expend $n_e$ [defined by (\ref{ne})] as
\begin{eqnarray}
\hspace*{-8mm}&&n_e=\left(\frac{1}{\gamma_e}\right)\phi + \left(\frac{\gamma_2}{2!\gamma_e^2}\right)\phi^2 
             +\left(\frac{\gamma_2\gamma_3}{3!\gamma_e^3}\right)\phi^3+\cdot\cdot\cdot,
\label{ne-exp}
\end{eqnarray}
where $\gamma_2=2 - \gamma_e$ and $\gamma_3=3 - 2\gamma_e$. The substitution of $\gamma_e=1$ into (\ref{ne-exp}) leads to
\begin{eqnarray}
&&n_e=1+\phi+\frac{\phi^2}{2!}+\frac{\phi^3}{3!}+\cdot\cdot\cdot=\exp(\phi).
\label{ne-Boltzmann}
\end{eqnarray}
Thus, (\ref{ne}) [by rewriting it  in the form of (\ref{ne-exp})] is also valid for $\gamma_e=1$ corresponding to  BDES, $n_e=\exp(\phi)$ with $\phi=e\Phi/k_BT_e$ and  $\sigma_l=T_i/Z_lT_e$.

On the other hand, substituting $P_{lt}$ [which can be obtained from (\ref{P6})] into (\ref{be4}),  the basic equations (\ref{be1}) and (\ref{be4}) for the NDLNS, and the Poisson equation (\ref{be5}) can be  reduced to their normalized form as
\begin{eqnarray}
&&\frac{\partial n_l}{\partial t} +\frac{\partial}{\partial x}(n_lu_l) = 0,
\label{be6}\\
&&\frac{\partial u_l}{\partial t} +u_l\frac{\partial u_l}{\partial x}=-\frac{\partial\phi}{\partial x}-\frac{\sigma_l}{n_l}\frac{\partial n^{\gamma_l}}{\partial x},
\label{be7}\\
&&\frac{{\partial}^2\phi}{\partial x^2}=(1+\mu)n_e-n_l-\mu,
\label{be8}
\end{eqnarray}
where  $n_l=N_l/N_{l0}$, $u_l=U_l/C_q$, $\phi=e\Phi/{\cal E}_{eT}$,  $t=T\omega_{pl}$,  $x=X/\Lambda_q$, and $\sigma_l=T_i/Z_l{\cal E}_{eT}$.  

\section{NA Solitary Waves}
To study arbitrary amplitude NA SWs by the pseudo-potential approach \cite{Bernstein57,Cairns95},  we first assume that all dependent variables in (\ref{be6})-(\ref{be8}) depend on a single variable $\xi= x-{\cal M}t$ (with ${\cal M}$ being the Mach number), and that $\partial /\partial t \rightarrow 0$ (corresponding to stationary NA SWs).  These assumptions allow us to express (\ref{be6})-(\ref{be8}) as
\begin{eqnarray}
&&{\cal M}\frac{dn_l}{d\xi}-\frac{d}{d\xi}(n_lu_l) = 0,
\label{sw1}\\
&&{\cal M}\frac{du_l}{d\xi}-u_l\frac{d u_l}{d\xi}=\frac{d\phi}{d\xi}+\frac{\sigma_l}{n_l}\frac{dn_l^{\gamma_l}}{d\xi},
\label{sw2}\\
&&\frac{d^2\phi}{d\xi^2}=(1+\mu)n_e-n_l-\mu.
\label{sw3}
\end{eqnarray}
To substitute $u_l$ [to be obtained from (\ref{sw1})]  into (\ref{sw2}), we express  (\ref{sw1})  and (\ref{sw2}) as
\begin{eqnarray}
\hspace*{-6mm}&&\frac{d}{d\xi}[{\cal M}n_l-n_lu_l] = 0,
\label{sw4}\\
\hspace*{-6mm}&&\frac{d}{d\xi}\left[{\cal M}u_l-\frac{1}{2}u_l^2-\phi-\left(\frac{\sigma_l\gamma_l}{\gamma_l-1}\right)n_l^{(\gamma_l-1)}\right]=0,
\label{sw5}
\end{eqnarray}
The integration of (\ref{sw4}) and (\ref{sw5}) gives rise to
\begin{eqnarray}
&&u_l={\cal M}\left(1-\frac{1}{n_l}\right),
\label{sw6}\\
&&2{\cal M}u_l-{u_l}^2-2\phi-\gamma_\sigma [n_l^{(\gamma_l-1)}-1]=0,
\label{sw7}
\end{eqnarray}
where $\gamma_\sigma=2\sigma_l\gamma_l/(\gamma_l-1)$ and the integration constants are determined by assuming the equilibrium state of TDP system (viz.  $u_l=0$, $\phi=0$, and $n_l=1$).  Now, substituting (\ref{sw6})  into (\ref{sw7}), we obtain
\begin{eqnarray}
&&\gamma_\sigma n_l^{(\gamma_l+1)}-({\cal M}^2+\gamma_\sigma -2\phi)n_l^2 + {\cal M}^2=0,
\label{sw8}
\end{eqnarray}
which is valid for arbitrary  value of $\gamma_e$ (except $\gamma_l=1$).  It is valid for cold NDLNS ($\sigma_l=0$),  and non-degenerate adiabatic NDLNS ($\gamma_l=3$).  We have neglected the effect of the light nucleus degeneracy in our present work since the degeneracy in light nucleus species is at least $ (m_e/m_l)^2$ times lower than that in electron species \cite{Shukla2011b,Mamun2018,MAS2016,MAS2017}.  The cold NDLNS ($\sigma_l=0$) gives (\ref{sw8}) to  
\begin{eqnarray}
&&n_l=\frac{1}{\sqrt{1-\frac{2\phi}{{\cal M}^2}}}.
\label{nl0}
\end{eqnarray} 
This means that we can consider $\sigma_l=0$, but cannot consider  $\gamma_l=1$ with $\sigma_l\ne0$ to study the effect of the NDLNS temperature on the basic features of arbitrary amplitude SWs by the pseudo-potential approach.

Now, for warm adiabatic NDLNS ($\sigma_l \ne 0$ and $\gamma_l=3$)  we can express $n_l$ from  (\ref{sw8}) as
\begin{eqnarray}
\hspace*{-8mm}&&3\gamma_l n_l^4-({\cal M}^2+3\sigma_l -2\phi)n_l^2 + {\cal M}^2=0,
\label{sw10}
\end{eqnarray} 
The solution of (\ref{sw10}), which is a quadratic equation for $n_l^2$, for $n_l$ is given by
\begin{eqnarray}
\hspace*{-6mm}&&n_l=\left[\frac{1}{6\sigma_l}\left(\Phi_0-\sqrt{\Phi_0^2-12\sigma_l {\cal M}^2}\right)\right]^{\frac{1}{2}},
 \label{nl1}
\end{eqnarray}
where
$\Phi_0={\cal M}^2+3\sigma_l-2\phi$.

Now,  multiplying both side of (\ref{sw5})  by
$d\phi/d\xi$ and integrating with respect to $\xi$, we obtain
\begin{eqnarray}
&&\frac{1}{2}\left(\frac{d\phi}{d\xi}\right)^2+V(\phi)=0,
\label{EI}
\end{eqnarray}
where 
\begin{eqnarray}
V(\phi)= -\int [(1+\mu)n_e -n_l-\mu]d\phi,
\label{PP}
\end{eqnarray}
in which $n_e$ is  given by (\ref{ne}), which is valid $\gamma=5/3$ (non-relativistically TDES) and  $\gamma=4/3$
(ultra-relativistically TDES) or (\ref{ne-exp}), which is valid for $\gamma=1$ (BDES),
 and  $n_l$ is given by (\ref{nl0}), which valid for $\sigma_l=0$ or (\ref{nl1}), which is valid for  
$\gamma_l=3$ (adiabatic NDLNS).  We note that (\ref{EI}) is an energy integral for  a  pseudo-particle of 
unit mass with $\xi$ as pseudo-time, $\phi$ as pseudo-position, and $V(\phi)$ as pseudo-potential. 
Therefore, from the analysis of $V(\phi)$, one can not only find the conditions for the formation of the NA SWs, 
but also can study their basic features.  We now consider the  following two situations of NDLNS number density $n_l$, 
and  investigate the basic features of the NA SWs.
\subsection{Cold NDLNS ($\sigma_l=0$)} 
The cold NDLNS ($\sigma_l=0$) is valid for $\omega/k\gg (k_BT_{l0} /m_l)^{1/2}$, which is usual for the NA waves and plasma system under consideration.  We, thus, first substitute  (\ref{ne})  and  (\ref{nl0})  into (\ref{PP}), and express (\ref{PP}) as
\begin{equation}
V(\phi)=C_0-\delta\left(1+\frac{\phi}{\gamma_e^\prime}\right)^{\gamma_e^\prime}-{\cal M}^2\sqrt{1-\frac{2\phi}{{\cal M}^2}}+\mu\phi,
\label{PP1}
\end{equation}
where  $C_0=1+\mu+M^2$ is the integration constant, and is chosen  in such a way that $V(0)=0$, $\delta=1+\mu$,  and 
$\gamma_e^\prime=\gamma_e/(\gamma_e-1)$.

\begin{figure}[htb]
 \includegraphics[width=0.48\textwidth]{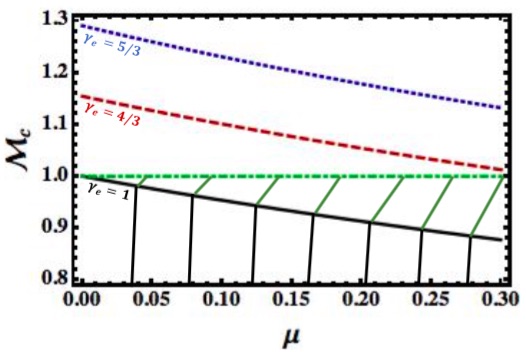} 
\caption{The variation of ${\cal M}_c$ with $\mu$ for  $\gamma_e=1$ (solid curve), $\gamma_e=5/3$ (dotted curve), and $\gamma_e=4/3$ (dashed curve). The dot-dashed curve represents ${\cal M}_c=1$.}
\label{AAM-f1-Mc}
\end{figure}

It is usual  for almost all non-degenerate and degenerate plasma situations that ${\cal E}_{eT}>{\cal E}{ed},~{\cal E}_{et}$,  and  $|\phi|<1$ since  
${\cal E}_{eT}={\cal E}{ed}+{\cal E}_{et}$,  $\phi=\Phi/{\cal E}_{eT}$, 
and ${\cal E}_{et}\simeq k_BT_{e0}$.  Thus, $V(\phi)$  [defined by (\ref{PP1})] can be expanded as
\begin{equation}
V(\phi)=C_2\phi^2+C_3\phi^3+\cdot \cdot \cdot,
\label{PPEXP1}
\end{equation}
where
\begin{eqnarray}
&&C_2=\frac{1}{2!}\left[\frac{1}{{\cal M}^2}-\frac{1}{\gamma_e}(1+\mu)\right],
\label{C2}\\
&&C_3= \frac{1}{3!}\left[\frac{3}{{\cal M}^4}-\frac{1}{\gamma_e^2}(2-\gamma_e)(1+\mu)\right].
\label{C3}
\end{eqnarray}
It is clear from (\ref{PPEXP1}) that the constant $C_0$,  and the coefficient of $\phi$ in the expansion of $V(\phi)$ vanish because of the choice of the  integration constant,  and because  of the equilibrium charge neutrality condition, respectively.  Thus, the NA SWs exist  if and only if  $[d^2V/d\phi^2]_{\phi=0}<0$ so that the fixed point at the origin is unstable \cite{Cairns95}, and  $[d^3V/d\phi^3]_{\phi=0}>~(<)0$ for the existence of the NA SWs with $\phi>0$ ($\phi<0$) \cite{Cairns95}.  These imply that the NA SWs exist if $C_2<0$,  i.e.  if  ${\cal M}>{\cal M}_c$, where 
${\cal M}_c$ is the critical Mach number (the minimum value of the Mach number above which the NA SWs exist), and is given by
\begin{eqnarray}
{\cal M}_c=\sqrt{\frac{\gamma_e}{1+\mu}}.
\label{Mc}
\end{eqnarray}
On the other hand, the NA SWs exist with $\phi>0$ ($\phi<0$) if $C_3 ({\cal M}={\cal M}_c)>0~(<0)$, where  $C_3 ({\cal M}={\cal M}_c)$ is
\begin{eqnarray}
C_3 ({\cal M}={\cal M}_c)=\left(\frac{1+\mu}{3!\gamma_e^2}\right)(1+\gamma_e+3\mu),
\label{C3-Mc}
\end{eqnarray}
which  implies that  $C_3 ({\cal M}={\cal M}_c)>0$ since $\mu\ge 0$ and $\gamma_e\ge 1$,  and that the NA SWs only with $\phi>0$ exist for any possible values of $\mu$ and  $\gamma_e$ (so, from now `SWs' will be used to mean`SWs with  $\phi>0$').  We have examined the variation of ${\cal M}_c$  with $\mu$ for 
$\gamma_e=1$ (BDES),  $\gamma_e=5/3$  (non-relativistically DES), and $\gamma_e=4/3$  (ultra-relativistically DES) as shown, respectively, in solid, dotted, and dashed curves of figure \ref{AAM-f1-Mc}. 
This figure clearly indicates that for  BDES, the subsonic NA SWs exist for those values of $\mu$, which are in between solid and dot-dashed curves,  and that above the dot-dashed curve, there exist the supersonic NA SWs.  On the other hand,  for realistic values of $\mu$  (e.g.  $0<\mu<0.3$),  the non-relativistically and ultra-relativistically DES are in against for the formation of subsonic NA SWs, but are in favor of the formation of supersonic NA SWs. 
   
We now  study the basic features of small amplitude NA SWs by considering the approximation
$V(\phi)=C_2\phi ^2+C_3\phi^3$. This approximation along with the condition $V(\phi_m)=0$ (where 
$\phi_m~(\ne 0)$ is the amplitude of the solitary waves) reduces the SW solution of (\ref{EI}) to
\begin{eqnarray}
\phi=\left(-\frac{C_2}{C_3}\right){\rm
sech^2}\left(\sqrt{-\frac{C_2}{2}}\xi\right), 
\label{sas1}
\end{eqnarray}
which has been derived in Appendix A.  

\begin{figure}[htb]
\includegraphics[width=0.48\textwidth]{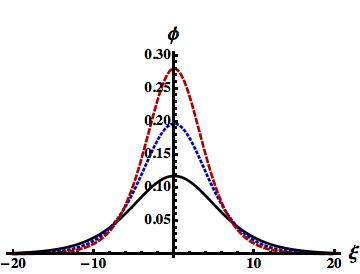} 
\caption{The small amplitude subsonic NA SWs for $\gamma_e=1$,  ${\cal M}=0.99$,
 $\mu=0.1$ (solid curve),  $\mu=0.15$  (dotted curve), and $\mu=0.2$ (dashed curve).} 
\label{AAM-f2-SAS}
\end{figure}

\begin{figure}[htb]
\includegraphics[width=0.48\textwidth]{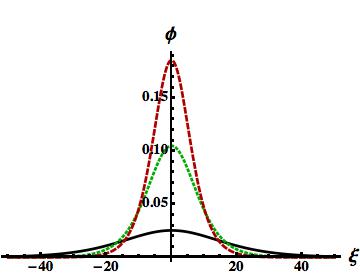} 
\caption{The small amplitude supersonic (${\cal M}=1.24$) NA SWs for $\gamma_e=5/3$,  $\mu=0.1$,
 (solid curve), $\mu=0.15$ (dotted curve), and $\mu=0.2$ (dashed curve).}
\label{AAM-f3-SAS}
\end{figure}

\begin{figure}[htb]
\includegraphics[width=0.48\textwidth]{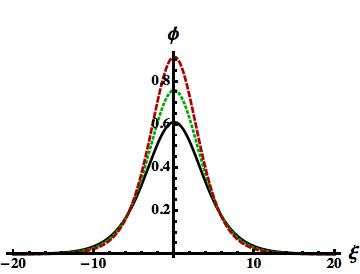} 
\caption{The small amplitude supersonic NA SWs for $\gamma_e=4/3$, ${\cal M}=1.24$,  $\mu=0.1$,
 (solid curve), $\mu=0.15$ (dotted curve), and $\mu=0.2$ (dashed curve).}
\label{AAM-f4-SAS}
\end{figure}

\begin{figure}[htb]
 \includegraphics[width=0.48\textwidth]{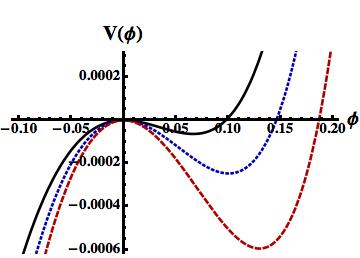} 
\caption{The potential wells  formed in $+\phi$-axis (corresponding to the formation of subsonic NA SWs) for $\gamma_e=1$,  
${\cal M}=0.99$, $\mu=0.1$ (solid curve),  $\mu=0.15$  (dotted curve), and $\mu=0.2$ (dashed curve).}
\label{AAM-f5-AAS}
\end{figure}

\begin{figure}[htb]
\includegraphics[width=0.48\textwidth]{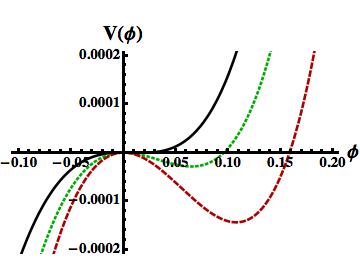} 
\caption{The potential  wells formed in  $+\phi$ axes (corresponding to the formation of supersonic 
NA SWs)  for $\gamma_e=5/3$, ${\cal M}=1.24$, $\mu=0.1$ (solid curve), $\mu=0.15$ (dotted curve), and $\mu=0.2$ (dashed curve).}
\label{AAM-f6-AAS}
\end{figure}

\begin{figure}[htb]
\includegraphics[width=0.48\textwidth]{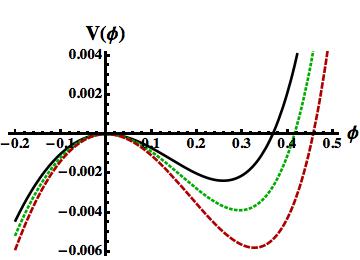} 
\caption{The potential  wells formed in $+\phi$ axes [corresponding to the formation of supersonic 
NA SWs) for $\gamma_e=4/3$, ${\cal M}=1.24$,  $\mu=0.1$ (solid curve), $\mu=0.15$ (dotted curve), and $\mu=0.2$ (dashed curve).}
\label{AAM-f7-AAS}
\end{figure}

We have graphically represented  (\ref{sas1}) to observe  the basic features of small amplitude subsonic (${\cal M}=0.99$)  NA SWs for $\gamma_e=1$ (BDES),  $\gamma_e=5/3$ (non-relativistically DES), and$\gamma_e=4/3$  (ultra-relativistically DES) for different values of $\mu$ (viz. $\mu=0.1$, $\mu=0.15$, and 
$\mu=0.2$). The results are displayed in figures \ref{AAM-f2-SAS}$-$\ref{AAM-f4-SAS}.  We have also reexamined these basic features of these subsonic and supersonic NA SWs by the direct analysis of pseudo-potential $V(\phi)$ defined by (\ref{PP1}) for the same set of plasma parameters. The results are displayed in figures \ref{AAM-f5-AAS}$-$\ref{AAM-f7-AAS}. 

It is obvious from figures \ref{AAM-f2-SAS}$-$\ref{AAM-f7-AAS} that (i) the presence of  SHNS  supports the existence of small amplitude subsonic NA SWs; (ii) the amplitude (width) of the subsonic NA SWs  increases (decreases) with the increase in number density (represented by $\mu$) of the SHNS;  (iii) the effects of non-relativistically ($\gamma_e=5/3$) and ultra-relativistically ($\gamma_e=4/3$) DES are in against the formation of the subsonic NA SWs, and thus, give rise to the formation of the supersonic NA SWs; (iv) the amplitude of the supersonic NA SWs  in no-relativistically DES ($\gamma_e=5/3$) is much  smaller than that in 
ultra-relativistically DES ($\gamma_e=4/3$), but is much larger than that in BDES ($\gamma_e=1$);  (iv) the width of the supersonic NA SWs in non-relativistically DES ($\gamma_e=5/3$) is much wider than that in ultra-relativistically DES ($\gamma_e=4/3$); (v) the small amplitude approximation provides almost the same  results as the direct analysis of the pseudo-potential $V(\phi)$ [defined by (\ref{PP1})] does.

The amplitude and  the width of the NA SWs are also visualized  from figures \ref{AAM-f5-AAS}$-$\ref{AAM-f7-AAS}. The potential wells in  figures \ref{AAM-f5-AAS}$-$\ref{AAM-f7-AAS} indicate the amplitude $\phi_m$ (value of $\phi$ at the point where the $V(\phi)$ vis. $\phi$ curve  crosses the $\phi$-axis), and the width ${\cal W}$ [defined as ${\cal W}=\phi_m/\sqrt{|V_m|}$, where $|V_m|$ is the  maximum value of $V(\phi)$ in the potential wells.  Thus, the  figures \ref{AAM-f5-AAS}$-$\ref{AAM-f7-AAS} indicate that the amplitude (with) of both subsonic and supersonic  NA increases (decrease) with the rise of $\mu$, and that their amplitude (width) of both subsonic and supersonic  NA decrease (increase) with the rise of $\gamma_e$, since in comparison with an increase in $\phi_m$,  a very slight increase/decrease in $|V_m|$ causes a very significant decrease/increase in ${\cal W}$. The same results have already been obtained from the analysis of the SW solution (\ref{sas1}), which is valid for the small, but finite amplitude subsonic and supersonic NA SWs.  
\subsection {Warm NDLNS ($\sigma_l>0$)} 
We now consider warm adiabatic NDLNS of number density defined by  (\ref{nl1}).
The latter is valid when $P_{dl}\ll P_{tl}$, which is valid not only for hot hot white dwarfs  \cite{Dufour08,Dufour11,Werner15,Werner19,Koester20},  but also for many space \cite{Rosenberg95,Havnes96,Tsintikidis96,Gelinas98} and laboratory  \cite{Fortov96,Fortov98,Mohideen98} plasma situations.
Thus,  substituting  (\ref{ne}) and (\ref{nl1})  into (\ref{PP}), and following the same procedure as adopted before, we can express the  pseudo-potential $V(\phi)$  [in the  energy integral defined by (\ref{EI})]  as
\begin{eqnarray}
\hspace*{-10mm}&&V(\phi)=C_0^\sigma+\mu\phi-(1+\mu)\left[1+\left(\frac{\gamma_e-1}{\gamma_e}\right)\phi\right]^{\frac{\gamma_e}{\gamma_e-1}} \nonumber\\
\hspace*{-10mm}&&~~~~~~~~~~~~~~-\frac{\sqrt{2}}{3\sqrt{3\sigma_l}}\left(\sqrt{\Phi_0-\Phi_1}\right)\left(\Phi_0+\frac{1}{2} \Phi_1\right), 
\label{PP2}
\end{eqnarray}
where  $C_0^\sigma=1 +\mu +\sigma_l+{\cal M}^2$ is the integration constant chosen in such a way that $V(\phi)=0$ at $\phi=0$, $\Phi_0={\cal M}^2+3\sigma_l-2\phi$, and $\Phi_1=\sqrt{\Phi_0^2-12\sigma_l {\cal M}^2}$.  The charge neutrality condition at equilibrium (viz. $n_e=1$ and $n_l=1$) leads to  
$[dV/d\phi ]_{\phi=0}=0$.  To have the NA SW solution of (\ref{EI}), its pseudo-potential  $V(\phi)$  [defined (\ref{PP2})]  must have an unstable fixed point \cite{Cairns95} at the origin ($\phi=0$),  i.e.   $[d^2V/d\phi^2]_{\phi=0}<0$, and at the same time (i.e. satisfying this condition) if 
$[d^3V/d\phi^3]_{\phi=0}>0~(<0)$, the NA SWs with $\phi>0~ (\phi<0)$ exist \cite{Cairns95}.

\begin{figure}[htb]
 \includegraphics[width=0.48\textwidth]{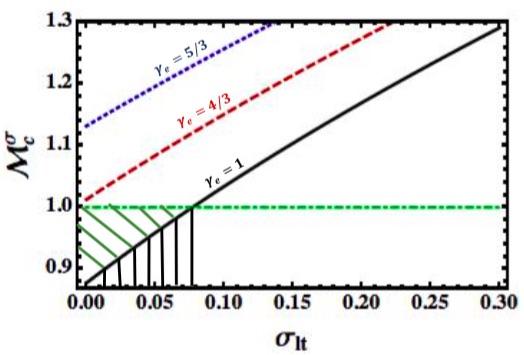} 
\caption{The variation of ${\cal M}_c^\sigma$ with $\sigma_l$ for $\mu=0.2$, $\gamma_e=1$ (solid curve), $\gamma_e=5/3$ (dotted curve), and $\gamma_e=4/3$ (dashed curve). The dot-dashed curve represents ${\cal M}_c^\sigma=1$.} 
\label{AAM-f8-Mc}
\end{figure}

To find the conditions for the existence  of the NA SWs analytically, we expand  $V(\phi)$ defined by (\ref{PP2}) as  
\begin{eqnarray}
V(\phi)= C_2^\sigma\phi^2+ C_3^\sigma\phi^3 + \cdot \cdot \cdot, 
\label{PPEXP2}
\end{eqnarray}
where
\begin{eqnarray}  
\hspace*{-6mm}&&C_2^\sigma=\frac{1}{2!}\left[\frac{1}{{\cal M}^2-3\sigma_{lt}} -\frac{1}{\gamma_e}(1+\mu)\right],
\label{C2s}\\
\hspace*{-6mm}&&C_3^\sigma=\frac{1}{3!}\left[\frac{3({\cal M}^2+\sigma_{lt})}{({\cal M}^2-3\sigma_{lt})^3}-\frac{1}{\gamma_e^2}(2-\gamma_e)(1+\mu)\right].
\label{C3s}
\end{eqnarray}
The coefficient of $\phi^2$ (viz. $C_2^\sigma$)  indicates from $[d^2V/d\phi^2]_{\phi=0}<0$)  that the SW solution of (\ref{EI}) with (\ref{PP2}) exists if and only if $C_2^\sigma<0$. Thus, the NA SWs exist if ${\cal M}>{\cal M}_c^\sigma$, where  
${\cal M}_c^\sigma$ is given by
\begin{eqnarray}
{\cal M}_c^\sigma=\sqrt{\frac{\gamma_e}{1+\mu}+3\sigma_l}.
\label{Mcs}
\end{eqnarray}
On the other hand, the NA SWs exist with $\phi>0$ ($\phi<0$) if $C_3 ({\cal M}={\cal M}_c)>0~(<0)$, where  $C_3 ({\cal M}={\cal M}_c)$ is
\begin{eqnarray}
C_3 ({\cal M}={\cal M}_c^\sigma)=\left(\frac{\delta}{3!\gamma_e^2}\right)\left(\gamma_0+3\mu+\frac{12}{\gamma_e}\sigma_l\delta^2\right),
\label{C3-Mcs}
\end{eqnarray}
where $\delta=1+\mu$ and $\gamma_0=1+\gamma_e$.  Equation (\ref{C3-Mcs}) implies that  $C_3 ({\cal M}={\cal M}_c)>0$ (since $\mu\ge 0$, 
$\gamma_e\ge 1$, and $\sigma_l\ge 0$) that the NA SWs only with $\phi>0$ exist for all possible values of $\mu$, $\gamma_e$, and $\sigma_l$. 
It is clear from  (\ref{Mcs}) that ${\cal M}_c^\sigma={\cal M}_c$ for $\sigma_l=0$. We have graphically shown the variation of ${\cal M}_c^\sigma$ 
with $\sigma_l$ for $\mu=0.2$. This is shown in figure \ref{AAM-f8-Mc}.  
which shows that (i) ${\cal M}_c^\sigma$ increases with $\sigma_l$;  (ii) the increase in $\sigma_l$ increases the minimum 
value of $\mu$ for which  the subsonic NA SWs exist; (iii) the existence of subsonic NA SWs for 
$\mu=0.3$  and a short range of  $\sigma_l$ is  shown in between solid and dot-dashed curves, and that above  the dot-dashed curve,  there exist the supersonic NA SWs;  (iv) for realistic values of $\mu$ (e.g.  $\mu<0.3$), the non-relativistically and ultra-relativistically electron degeneracies as well as light nucleus temperature are in against the formation of subsonic NA SWs, but are in favor of the formation of supersonic NA SWs. 

We again study small amplitude NA SWs for which
$V(\phi)=C_2^\sigma\phi^2+ C_3^\sigma\phi^3$ holds good. This approximation along with the condition $V(\phi_m)=0$ allows us to write the SW solution of (\ref{EI})  as
\begin{eqnarray}
\phi=\left(-\frac{C_2^\sigma}{C_3^\sigma}\right){\rm
sech}^2\left(\sqrt{-\frac{C_2^\sigma}{2}}\xi\right),
\label{sas2}
\end{eqnarray} 
which is also derived in Appendix A.

\begin{figure}[htb]
\includegraphics[width=0.48\textwidth]{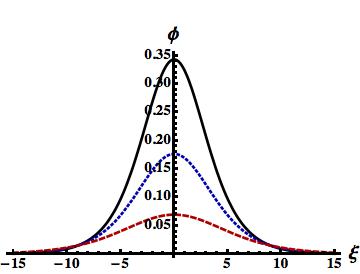} 
\caption{The small amplitude subsonic (${\cal M}=0.99$) NA SWs for $\gamma_e=1$,  
$\mu=0.3$,  $\sigma_l=0.01$ (solid curve), $\sigma_l=0.02$ (dotted curve), and $\sigma_l=0.03$ (dashed curve).} 
\label{AAM-f9-SAS}
\end{figure}

\begin{figure}[htb]
\includegraphics[width=0.48\textwidth]{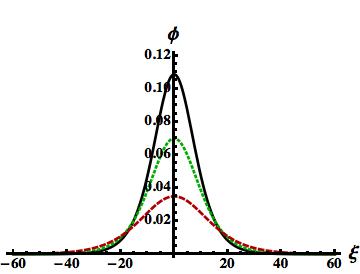} 
\caption{The small amplitude supersonic (${\cal M}=1.28$) NA SWs for $\gamma_e=5/3$, $\sigma_l=0.01$ (solid curve), 
$\sigma_l=0.02$ (dotted curve), and $\sigma_l=0.03$ (dashed curve).}
\label{AAM-f10-SAS}
\end{figure}

\begin{figure}[htb]
\includegraphics[width=0.48\textwidth]{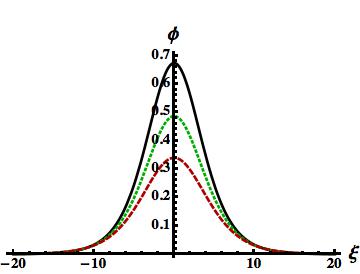} 
\caption{The small amplitude supersonic (${\cal M}=1.28$) NA SWs for $\gamma_e=4/3$, $\sigma_l=0.02$ (solid curve), 
$\sigma_l=0.04$ (dotted curve), and $\sigma_l=0.06$ (dashed curve).}
\label{AAM-f11-SAS}
\end{figure}

\begin{figure}[htb]
 \includegraphics[width=0.48\textwidth]{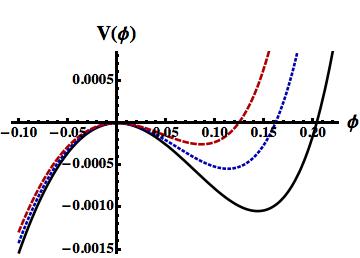} 
\caption{The potential wells formed in $+\phi$-axis [corresponding to the formation of subsonic 
(${\cal M}=0.99$) NA SWs for $\gamma_e=1$,  $\mu=0.3$,  $\sigma_l=0.01$ (solid curve), $\sigma_l=0.02$ (dotted curve), and $\sigma_l=0.03$ (dashed curve).}
\label{AAM-f12-AAS}
\end{figure}

\begin{figure}[htb]
\includegraphics[width=0.48\textwidth]{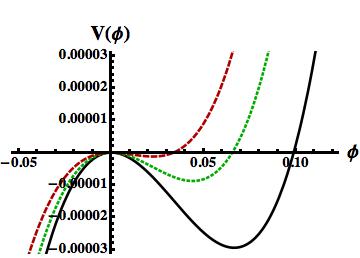} 
\caption{The potential wells  formed in $+\phi$-axis [corresponding to the formation of supersonic 
(${\cal M}=1.24$) NA SWs for $\gamma_e=5/3$, $\sigma_l=0.01$ (solid curve), 
$\sigma_l=0.02$ (dotted curve), and $\sigma_l=0.03$ (dashed curve).}
\label{AAM-f13-AAS}
\end{figure}

\begin{figure}[htb]
\includegraphics[width=0.48\textwidth]{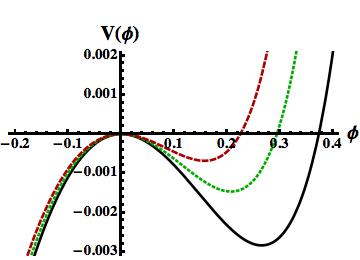} 
\caption{The potential wells  formed in $+\phi$-axis [corresponding to the formation of supersonic 
(${\cal M}=1.24$) NA  SWs for $\gamma_e=4/3$, $\sigma_l=0.02$ (solid curve), 
$\sigma_l=0.04$ (dotted curve), and $\sigma_l=0.06$ (dashed curve).}
\label{AAM-f14-AAS}
\end{figure}

To observe the effect of light nucleus temperature ($\sigma_l$)  on the basic features of small amplitude subsonic and supersonic NA SWs, we have  graphically analyzed (\ref{sas2}) for $\gamma_e=1$  (BDES), $\gamma_e=5/3$ (non-relativistically DES), and $\gamma_e=4/3$ (ultra-relativistically DES). The results are displayed in figures \ref{AAM-f9-SAS}$-$\ref{AAM-f11-SAS}.  We have also reexamined these basic features of these subsonic and supersonic NA SWs by the direct analysis of the pseudo-potential $V(\phi)$ [defined by (\ref{PP2})] for the same set of plasma parameters.  The results are displayed in figures 
\ref{AAM-f12-AAS}$-$\ref{AAM-f14-AAS}. 

It is obvious from figures \ref{AAM-f9-SAS}$-$\ref{AAM-f14-AAS} that (i) the effect of light nucleus temperature significantly reduces the possibility for the formation of the subsonic NA SWs, and  in the case of more light nucleus temperature  ($\sigma_l$), we need more number of the SHNS to have the existence  of 
subsonic NA SWs;  (ii) the combined effects of light nucleus temperature ($\sigma_l$), non-relativistically ($\gamma_e=5/3$), 
and ultra-relativistically ($\gamma_e=4/3$) DES are in against for the formation of subsonic NA SWs, and thus, give rise to the formation of the supersonic NA SWs;  (iii) the amplitude of the supersonic NA SWs  in non-relativistically DES ($\gamma_e=5/3$) is much smaller than that in ultra-relativistically DES ($\gamma_e=4/3$), but is much larger than that BDES ($\gamma_e=1$); (iv) the width of the supersonic NA SWs  in non-relativistically DES ($\gamma_e=5/3$) is much wider than that in ultra-relativistically DES ($\gamma_e=4/3$); (v)  the amplitude (width) of both subsonic and supersonic  NA SWs decreases (increases) with the rise of the light nucleus  temperature $T_{l0}$; (vii) the small amplitude approximation provides almost the same  results as the direct analysis of the pseudo-potential $V(\phi)$ [defined by (\ref{PP2})] does.
\section{Discussion}
We consider a fully ionized multi-nucleus plasma system containing thermally degenerate electron, cold/warm non-degenerate light nucleus, and low dense heavy 
nucleus species.  The basic features of thermal and degenerate pressure driven arbitrary amplitude subsonic and supersonic nucleus-acoustic solitary waves 
in such a plasma system have been investigated by the pseudo-potential approach, which is valid for arbitrary amplitude nucleus-acoustic solitary waves. 
The results, which have been found from this investigation, can be pinpointed as follows:
\begin{itemize} 
\item{The presence of  stationary heavy nucleus species ($\mu>0$) in electron-nucleus plasma supports the existence of subsonic nucleus-acoustic  solitary waves 
with $\phi>0$.  This is due to the fact that the phase speed ($\omega/k$) of the nucleus-acoustic waves decreases with the rise of the number density of the stationary 
heavy nucleus species.}

\item{It has been observed that in the case of higher light nucleus temperature ($T_l$), we need more number of stationary heavy nucleus species to have the existence
of the subsonic nucleus-acoustic solitary waves.}

\item{The effects of non-relativistically ($\gamma_e=5/3$) and ultra-relativistically ($\gamma_e=4/3$) degenerate electron species, and 
light nucleus temperature ($\sigma_l$) reduce the possibility for the formation of subsonic nucleus-acoustic solitary waves, and thus, give rise to the formation 
of the supersonic nucleus-acoustic solitary waves  with $\phi>0$. This is because that the phase speed ($\omega/k$) increases with the increase in value of 
the index $\gamma_e$ and light nucleus temperature (represented by $\sigma_l$).}

\item{The amplitude of the supersonic nucleus-acoustic solitary waves  in non-relativistically degenerate electron species ($\gamma_e=5/3$) is much smaller than that 
in ultra-relativistically degenerate electron species ($\gamma_e=4/3$), but is much larger than that in  Boltzmann distributed electron species ($\gamma_e=1$).}

\item{The amplitude (width) of the subsonic and supersonic nucleus-acoustic solitary waves  increases (decreases) with the rise of the number density of the heavy 
nucleus species, represented by $\mu$. The width of the supersonic nucleus-acoustic solitary waves  in non-relativistically degenerate electron species ($\gamma_e=5/3$) 
is much wider than that in ultra-relativistically degenerate electron species ($\gamma_e=4/3$).}

\item{The amplitude (width) of both subsonic and supersonic  nucleus-acoustic solitary waves  decreases (increases) with the rise of the light nucleus  temperature, represented by $\sigma_l$. This is due to the fact that the temperature of the light nucleus fluid  enhances the random motion of light nucleus, which causes to decrease the amplitude of the NA solitary structures.}

\item{The correctness of the results are verified by obtaining the same results from the analysis of analytical solitary wave solution of the energy integral [defined by (\ref{EI})  with the pseudo-potential $V(\phi)$ provided in (\ref{PP1}) and (\ref{PP2})] and the direct analysis of the pseudo-potential $V(\phi)$ [provided in (\ref{PP1}) and (\ref{PP2})].}
\end{itemize}

There are many hot white dwarfs \cite{Dufour08, Dufour11,Werner15,Werner19,Koester20}, where the electron thermal pressure can be comparable to or greater than its degenerate pressure,  and  where in addition to degenerate electron species, non-degenerate light and heavy nucleus species exist.  
On the other hand, non-degenerate electron species [defined by (\ref{ne-Boltzmann}) as an special case of $\gamma_e=1$], ions [identical to light nucleus species considered here, and defined by (\ref{nl1})], and positively charged  particle (impurity/dust) species [identical to stationary heavy nucleus species considered here] are observed in both space \cite{Rosenberg95,Havnes96,Tsintikidis96,Gelinas98} and laboratory \cite{Fortov96,Fortov98,Mohideen98}  plasma situations.  

Therefore, the thermally degenerate plasma model under consideration is so general that it can be applied not only 
in astrophysical degenerate plasma systems \cite{Dufour08, Dufour11,Werner15,Werner19,Koester20},  but also in many space \cite{Rosenberg95,Havnes96,Tsintikidis96,Gelinas98} and laboratory  \cite{Fortov96,Fortov98,Mohideen98} plasma systems.  It may 
be added here that to examine the effects of the dynamics of heavy nucleus species and non-relativistic degeneracy in light nucleus species 
on the nucleus-acoustic subsonic and supersonic solitary waves (investigated in the present work) may also be a problem of great importance  
for some other degenerate plasma systems, but beyond the scope of the present work. However, it is expected that the present work is useful in understanding the physics of localized electrostatic disturbances in a number of astrophysical \cite{Dufour08, Dufour11,Werner15,Werner19,Koester20}, space \cite{Rosenberg95,Havnes96,Tsintikidis96,Gelinas98}, and laboratory \cite{Fortov96,Fortov98,Mohideen98}.   
plasma systems.  
\appendix
\section{\bf SW solution of $ \frac{1}{2}\left(\frac{d\phi}{d\xi}\right)^2+V(\phi)=0$, where $V(\phi)=C_2\phi^2+C_3\phi^3$}
To obtain the solitary wave (SW) solution of this energy integral, two conditions must be satisfied. These are  (i) $\left[{d^2\phi}/{d\xi^2}\right]_{\phi=0}<0$ 
and (ii) $V(\phi_m)=0$. The condition (i) means that the point of  $V(\phi)$ vs. $\phi$ curve  at the origin 
$(0, 0)$ is unstable, which is  satisfied if $C_2<0$.  The condition (ii) is satisfied  if
$C_2+C_3\phi_m=0$, which gives rise to $C_2=-C_3 \phi_m$ or $\phi_m=-C_2/C_3$.
Now, substituting $C_2=-C_3\phi_m$ into the energy integral, we have
\begin{eqnarray}
           &~&\frac{1}{2}\left(\frac{d\phi}{d\xi}\right)^2=C_3\phi^2(\phi_m-\phi)\nonumber\\
           &\Rightarrow&\left(\frac{d\phi}{d\xi}\right)^2=2C_3\phi^2(\phi_m-\phi)\nonumber\\
           &\Rightarrow&\frac{d\phi}{d\xi}=\sqrt{2C_3}\phi\sqrt{\phi_m-\phi}\nonumber\\
           &\Rightarrow&\frac{d\phi}{\phi\sqrt{\phi_m-\phi}}=\sqrt{2C_3}d\xi.
 \label{A1}
\end{eqnarray}
To integrate (\ref{A1}), we let $\sqrt{\phi_m-\phi}=z$, which yields $\phi=\left(\phi_m-z^2\right)$ and  $d\phi=-2zdz$. 
These along with $C_3 \phi_m=-C_2$ reduce ({A1}) to
\begin{eqnarray}
\hspace*{-6mm}&~&-\frac{2dz}{\phi_m-z^2}=\sqrt{2C_3}d\xi\nonumber\\
 \hspace*{-6mm}&\Rightarrow&-\left[\frac{1}{\sqrt{\phi_m}+z}+ \frac{1}{\sqrt{\phi_m}-z}\right]dz=\sqrt{-2C_2}d\xi.   
\label{A2}
\end{eqnarray}
The  integration of  (\ref{A2})  gives rise to
\begin{eqnarray}
&~~&\log(\sqrt{\phi_m}-z)-\log(\sqrt{\phi_m}+z)=\sqrt{-2C_2}\xi +K_0\nonumber\\
&\Rightarrow&\log\left(\frac{\sqrt{\phi_m}-z}{\sqrt{\phi_m}+z}\right)= \sqrt{-2C_2}\xi +K_0\nonumber\\
&\Rightarrow&\frac{\sqrt{\phi_m}-z}{\sqrt{\phi_m}+z}=K_1 \exp\left(\sqrt{-2C_2}\xi\right), 
\label{A3}
\end{eqnarray}
where $K_0$  is the integration constant,  and $K_1=\exp(K_0)$ is another constant to be determined.  We have   
 $\phi=\phi_m$  and  $z=0$  at  $\xi=0$  for the solitary wave solution. Thus,  $K_1$ is determined from (\ref{A3}) as  $K_1=1$,  
 and (\ref{A3}) can be expressed as
\begin{eqnarray}
z=\sqrt{\phi_m }\left[\frac{[1-\exp(\sqrt{2C_3\phi_m}\xi)}{1+\exp(\sqrt{2C_3\phi_m}\xi)}\right].
\label{A4}
\end{eqnarray}
Therefore, (\ref{A4}) reduces $\phi=\left(\phi_m-z^2\right)$ to
\begin{eqnarray}
\phi&=&\phi_m\left[1-\left\{\frac{[1-\exp(\sqrt{2C_3\phi_m}\xi)}{1+\exp(\sqrt{2C_3\phi_m}\xi)}\right\}^2\right] \nonumber\\
       &=&\phi_m\left[\frac{2\exp\left(\sqrt{\frac{C_3}{2}\phi_m}\xi\right)}{1+\exp(2\sqrt{\frac{C_3}{2}\phi_m}\xi)}\right]^2 \nonumber\\
       &=&\phi_m\left[\frac{2}{\exp\left(\sqrt{\frac{C_3}{2}\phi_m}\xi\right)+\exp\left(-\sqrt{\frac{C_3}{2}\phi_m}\xi\right)}\right]^2\nonumber\\
       &=&\phi_m~{\rm sech^2}\left(\sqrt{\frac{C_3}{2}\phi_m}\xi\right).  
\label{A5}
\end{eqnarray}
We note that the  last few finial step of (\ref{A5}) are obtained by using the basic properties of hyperbolic functions.  
Thus, substituting $\phi_m=-C_2/C_3$ into (\ref{A5}), we finally obtain 
\begin{eqnarray}
\phi=\left(-\frac{C_2}{C_3}\right){\rm sech^2}\left(\sqrt{-\frac{C_2}{2}}\xi\right).  
\label{A6}
\end{eqnarray}
\acknowledgements
The author acknowledges the financial support of Jahangirnagar University  Project  funded by the  University Grants Commissions of Bangladesh.

\end{document}